\documentclass[12pt]{article}
\usepackage{graphicx}
\usepackage{amssymb}
\input epsf

\def\beq{\begin{equation}}
\def\eeq{\end{equation}}

\title{\bf Elliptic and triangular flows in dAu collisions at 200 GeV in the fusing color string model.}
\author{M.A. Braun$^a$, C. Pajares$^b$\\
$^a$ Dep. of High Energy physics,
 Saint-Petersburg State University, Russia\\
$^b$ Dep. of Particles, University of Santiago de Compostela, Spain}

\begin{document}
\maketitle
\begin{abstract}
In the color string picture with fusion and percolation
the elliptic and triangular flows are studied for p-Au and d-Au collisions at 200 GeV.
The ordering $v_n(d-Au)>v_n(p-Au)$ observed experimentally for central collisions is reproduced.
The calculated elliptic flow $v_2$ at central collisions agrees satisfactorily with the
data. The triangular flow $v_3$ is found to be greater than the experimental values, similar to the results
obtained in the approach based on the Color Glass Condensate initial conditions with subsequent
hydrodynamical evolution.

\end{abstract}

%%%%%%%%%%%%%%%%%%%%%%%%%%%%%%%%
\section{Introduction}
One of the most impressing discoveries at RHIC and  LHC is observation of
strong azimuthal correlations in nucleus-nucleus collisions
~\cite{ref1,ref2,ref3,alver1,adamczyk,ref7}.
 It can be characterized by the non-zero
flow coefficients $v_n$ governing the correlation function of the azimuthal distribution of
secondaries as
\beq
C(\phi)=A(1+\Big(1+2\sum_{n=1}v_n\cos(n\phi)\Big).
\label{eq0}
\eeq
This effect can be understood as the formation of the fireball in the overlap of the colliding nuclei
consisting of the strongly interacting hot quark-gluon plasma, which subsequently freezes, hadronises
and passes into the observed secondary hadrons. The dynamics of this transition seems to be well described in
the hydrodynamical approach, which relates the final spatial anisotropy to that of the initial state.

Later a similar anisotropy was observed also for collisions of smaller systems such as p-p, p-A, d-A and He-A.
~\cite{khachatryan,Chatrchyan,abelev,aad,adare,adare1,aidala}.
This of course has raised doubts about formation of a significantly big pieces of quark-gluon plasma in the interaction region
and the subsequent hydrodynamical evolution. However calculations made within specific models of the latter ~\cite{bozek,sonic,iebe}
and also with  initial conditions created by  gluon emission in the Color Glass Condensate effective theory ~\cite{mace,mace1,schenke}
seem to describe at least part of the experimental data quite satisfactorily. So the dynamic assumptions adopted for A-A collisions
seem to work also for smaller systems.

This circumstance was earlier observed in an alternative scenario for high-energy collisions, namely, the fusing color string picture.
Much simpler that the hydrodynamical approach with or without previous gluon emission in the QCD framework, it allowed to
describe in a satisfactory way the dependence of the spectra both on the transverse momentum and angle at various energies and for
various colliding particles ~\cite{review}.
In this scenario the dynamics for  small and big participants is qualitatively the same.
The colliding nucleons form strings as soon as they are close enough and the strings then emit the observed secondary particles.
The angular anisotropy in this scenario is the result of their quenching due to the presence of the gluon field  from the created strings.
So essentially it is a two-stage scenario as opposed to three-stage scenarios consisting first in formation of the set of interacting nucleons,
then building of the initial condition  ( e.g. emission of gluons)  and finally the hydrodynamical expansion. Correspondingly it carries
only one adjustable parameter - the universal quenching coefficient to be extracted  from some data.

In this note we apply this approach to the elliptic and triangular flows $v_2$ and $v_3$  for p- and d-Au collisions at 200 GeV, recently measured by PHENIX
~\cite{aidala}. It is of especial interest due to speculations that the enhanced number of primary sources in the case of d-Au collisions should lead to smaller values
of $v_2$ because the different sources (strings) are separated and do not communicate \cite{aidala}. To overcome this difficulty in
most recent calculations with the QCD  different scales were introduced which characterize the individual domains of gluon emission;
its saturation momentum and gluon resolution. Within this framework a good description of $v_2(p_T)$ was achieved ~\cite{mace1,schenke}. As to $v_3(P_T)$ the model
overshoots the data although preserves the general trend with $p_T$.

As we shall find the fusing string model allows to satisfactorily describe $v_2(p_T)$ without any additional assumptions. The triangular flow
$v_3(p_T)$ is found to lie  greater than the data, as in ~\cite{sonic} and ~\cite{mace1,schenke}. The latter authors tend to ascribe this
overshooting to the details of hadronization and rescattering, which may be more important for $v_3$, since it is wholly fluctuation driven unlike $v_2$.
We are working to introduce some of this effect into our model

Notice, that the glasma picture of CGC in many aspects is similar to the fusion color
string model and a correspondence can be established between
characteristic quantities of both approaches. The saturation momentum,
occupation number and  number of color flux
tubes in the glasma picture correspond to to the square root of the
string density, fraction of the total collision area
occupied by strings and effective number of clusters of
strings in the fusion of color strings approach respectively.
This gives rise
to the same dependence on the energy and centrality of the main
observables in both approaches ~\cite{dedeu}. Thus, it is not surprising
that we find similar results for the flows.

\section{Flow coefficients in the color string model}
The flow coefficients are obtained after averaging over events of the inclusive particle distribution
in the azimuthal angle for a single event
\beq
I^e(\phi)=
A^e\Big[1+2\sum_{n=1}\Big(a_n^e\cos n\phi+
b_n^e\sin n\phi\Big)\Big].
\label{ie}
\eeq
The flow coefficients are
\beq
v_n=\Big<\Big[(a_n^e)^2+(b_n^e)^2\Big]^{1/2}\Big>.
\label{vnexp1}
\eeq

In experimental observations one often uses instead of (\ref{vnexp1})
\beq
v_n\{2\}=\Big(\Big<(a_n^e)^2+(b_n^e)^2\Big>\Big)^{1/2}.
\label{vnexp2}
\eeq
which is somewhat greater than $v_n$ defined by (\ref{vnexp1}).

In the color string model the event is realized by a particular way of exchange of color strings
between the projectile and target. Different events possess  different number of strings located at different places
in the overlap of the colliding nuclei. The model was proposed a long time ago to describe
multiparticle production in the soft region. Its latest development and applications are described in the review paper ~\cite {review}.
Here we only briefly reproduce the main
points necessary to understand the technique.
The strings that can be visualized as drops of strong gluonic field are  assumed
to possess a certain finite dimension in
the transverse space.
Each string eventually breaks down in parts several times until its energy becomes of the order
of  GeV and it becomes an observed hadron. The number of strings
grows with energy and atomic number and finally strings begin to overlap and fuse giving rise to clusters with
more color and covering more space in the interaction area. At a certain
critical density clusters acquire the transverse dimensions
comparable to that of the
interaction area (percolation).

It may be assumed that strings
decay into particles ($q\bar{q}$ pairs) by the well-known  Schwinger mechanism for
pair creation in a strong
electromagnetic field.
\beq
P(p,\phi)=Ce^{-\frac{p_0^2}{T}},
\label{prob}
\eeq
where $p_0$ is the particle initial transverse momentum,
$T$ is the string tension (up to an irrelevant numerical coefficient) and
$C$ is the normalization factor.
To extend  validity of the distribution to
higher momenta one may use the idea that the string tension fluctuates, which
transforms the Gaussian distribution into the thermal one ~\cite{bialas,deupaj}:
\beq
P(p,\phi)=Ce^{-\frac{p_0}{t}}
\label{probb}
\eeq
with temperature $t=\sqrt{T/2}$.

The initial transverse momentum $p_0$ is thought to be
different from the
observed particle momentum $p$ because
the particle has to pass through the  fused string areas and emit gluons
on its way out. So in fact in Eq. (\ref{prob}) or (\ref{probb}) one has to consider $p_0$
as a function of $p$ and path length $l$ inside each string encountered on its way out:
$p_0=f(p,l(\phi))$ where $\phi$ is the azimuthal angle. It is this quenching that creates the final anisotropy
and leads to nonzero flow coefficients, due to anisotropy od string distribution.
To describe this quenching
we use the corresponding QED picture for a charged particle  moving
in the external electromagnetic field ~\cite{nikishov}.
This leads to
the quenching formula inside a string passed by the parton~\cite{ref14}
\beq
p_0(p,l)=p\Big(1+\kappa p^{-1/3}T^{2/3}l\Big)^3,
\label{quench1}
\eeq
Here $l$ is the length traveled by the parton through the gluon fields in the hadron
formed by color strings. Note that both $l$ and $T$ are different for different strings.
When the parton passes through many strings inside the hadron one should
sum different $T^{2/3}_1l_1+ T^{2/3}_2l_2+...$ over all of them.
For an event both $T_i$ and $l_i$ are uniquely determined by the geometry of the collision
and string fusion. The quenching coefficient $\kappa$ to be taken from the experimental
data. We adjusted $\kappa$  to
give the experimental value for the coefficient $v_2$
in mid-central Pb-Pb collisions at 5-13 TeV GeV, integrated over the
transverse momenta, which gives $\kappa=0.6$.

Remarkably, Eq. (\ref{quench1}) gives rise to a universal dependence of $v_2$ on the
product  $\epsilon p^{2/3}T^{1/3}l$, where $\epsilon$ is the eccentrity of the nuclear overlap.
This scaling ia well confirmed by the experimental data ~\cite{andres1,andres2}

\section{Calculations}

In the Monte-Carlo simulations
of p-Au collisions at the first step one locates the interacting nucleons of the target, which are those
whose distance from the projectile nucleon does not exceed two nucleon radii. For d-Au collisions one distributes the
two nucleons in the deuteron according to the probability given by the deuteron wave function and locates interacting nucleons
for both of them using the same criterium, For the deuteron wave function we take the Hulthen
wave function.
Then each pair of interacting nucleons is presented as a disk
of the typical proton radius 0.8 fm and with the matter distributed inside according to
the Gaussian density. The strings exchanged between them  are then formed
randomly distributed in the transverse
plane and with the probability proportional to the hadron matter. Their number was taken from
the previous calculations depending on the energy and centrality. Overlapping strings are assumed to fuse
and their tension was determined from the string scenario (see ~\cite{review}).
Finally for a given point and  direction of emission one determines the strings which the emitted
particle crosses on its trajectory and lengths inside each string to find the final factor
$\sum T_i^{2/3}l_i$ entering the quenching formula. This allows to find the total probability of emission
using (\ref{prob}) or (\ref{probb}) with (\ref{quench1}). As a result one obtains the distribution in the azimuthal angle
for an event and, after averaging, coefficients $v_n$.

Each calculation is actually done at fixed impact parameter $b$. To compare with the experimental data referring to most
central collisions one has to take into account that the relation of $b$ and centrality is not direct due to strong fluctuations of the
number of interacting nucleons and strings at fixed $b$ ~\cite{bozek,adare1}. To separate the contribution from a given  centrality
therefore on has to study the multiplicity $\mu$ at each $b$ and each run and select events in which $\mu$ lies within a certain
part of the total interval $\mu<\mu_{max}$ where $\mu_{max}$ is the largest multiplicity for all $b$ and runs. In such an approach the most central
collisions can be defined as such with $0.9\mu_{max}<\mu<\mu_{\max}$. Likewise  mid-central collisions can be defined by
$0.45\mu_{max}<\mu<0.55\mu_{max}$ and very peripheral collisions by $\mu<0.1\mu_{max}$.

\begin{figure}
\begin{center}
\includegraphics[width=7.2 cm]{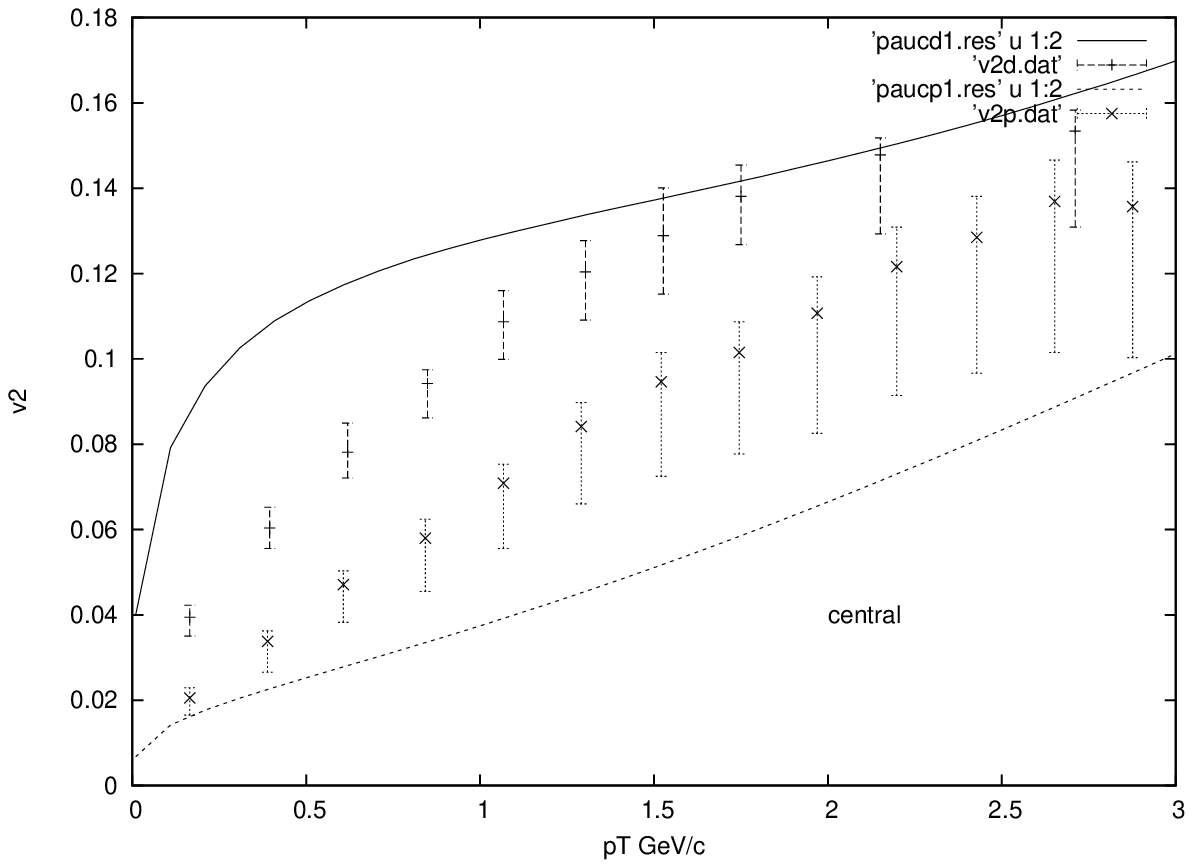}
\includegraphics[width=7.2 cm]{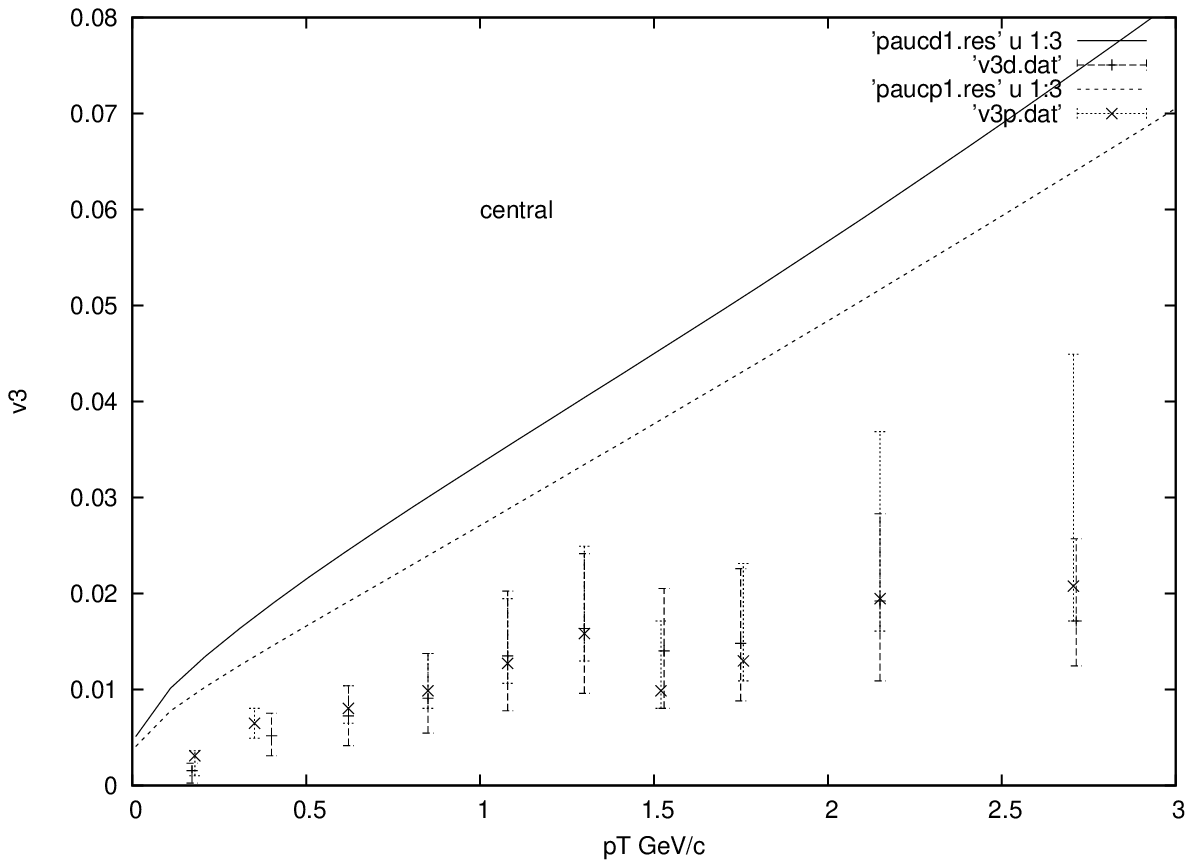}
\caption{The calculated flow coefficients $v_2$ (left panel) and $v_3$ (right panel) as function of transverse momenta $p_T$
for d-Au (upper curves) and p-Au  central collisions at 200 GeV.
Experimental data for $d-Au$ (upper points) and $p-Au$ collisions at 0-5\% centrality
are from ~\cite{aidala}.}
\label{fig1}
\end{center}
\end{figure}

\begin{figure}
\begin{center}
\includegraphics[width=7.2 cm]{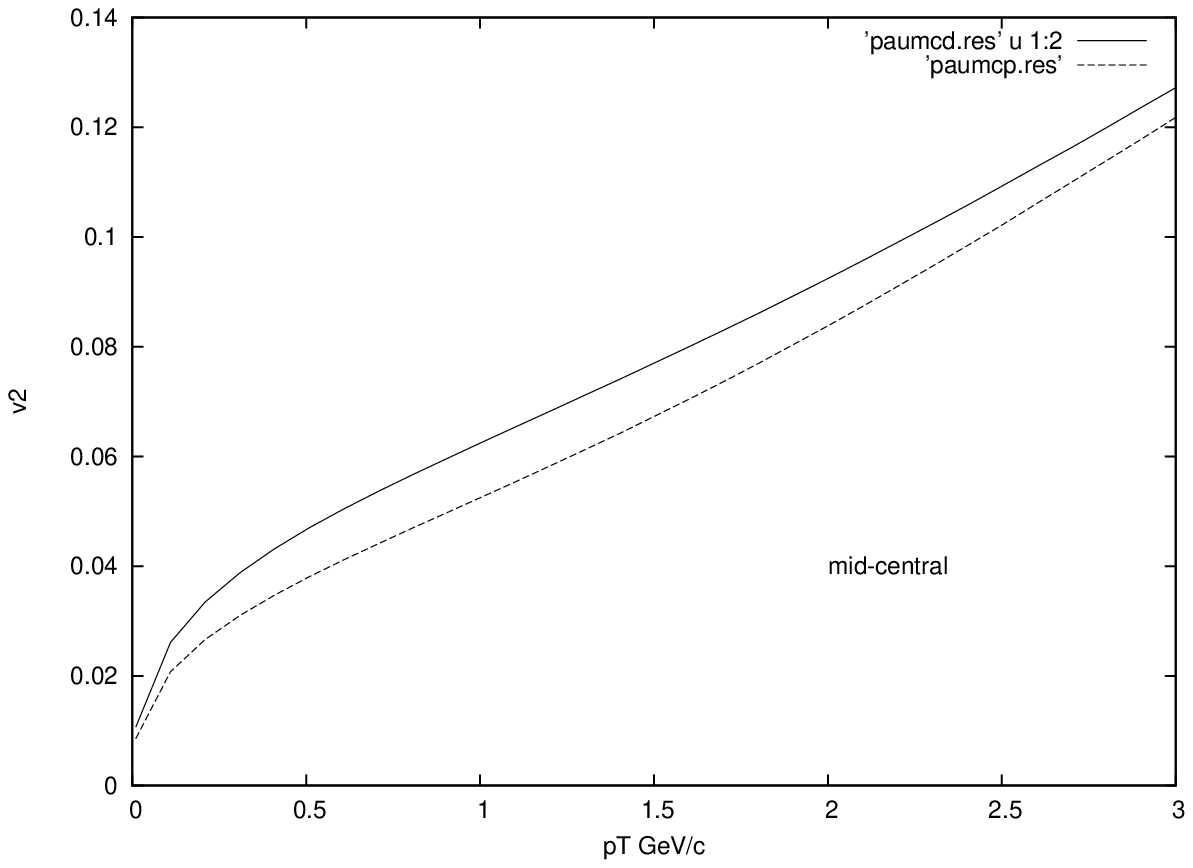}
\includegraphics[width=7.2 cm]{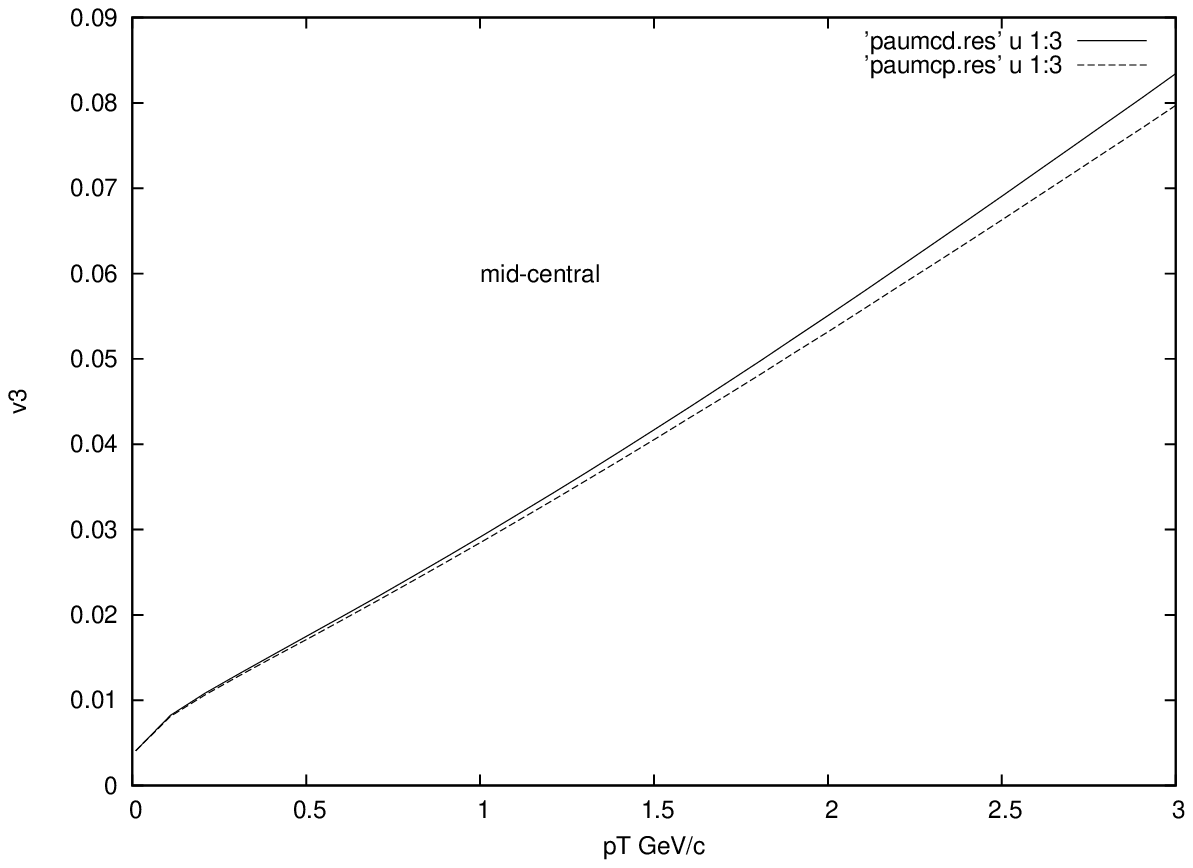}
\caption{The calculated flow coefficients $v_2$ (left panel) and $v_3$ (right panel) as function of transverse momenta $p_T$
for d-Au (upper curves) and p-Au  mid-central collisions at 200 GeV.}
\label{fig2}
\end{center}
\end{figure}

\begin{figure}
\begin{center}
\includegraphics[width=7.2 cm]{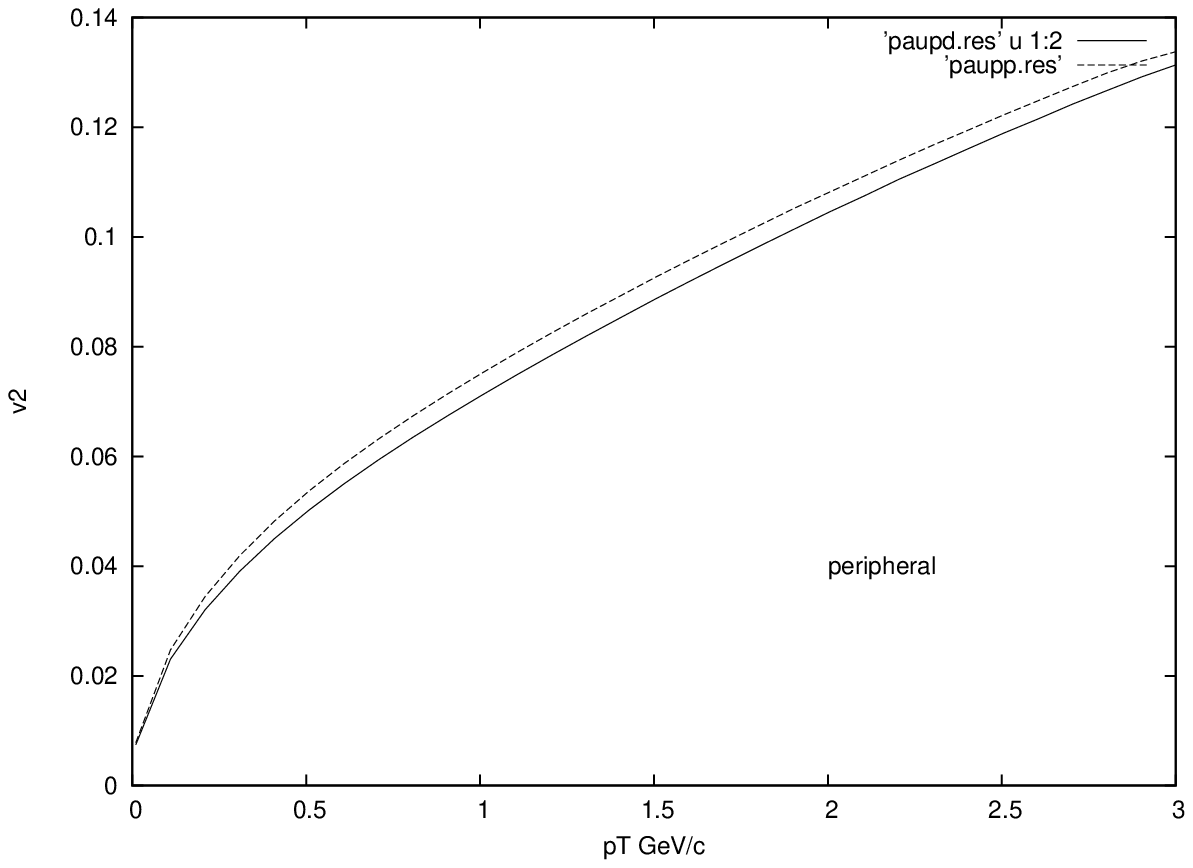}
\includegraphics[width=7.2 cm]{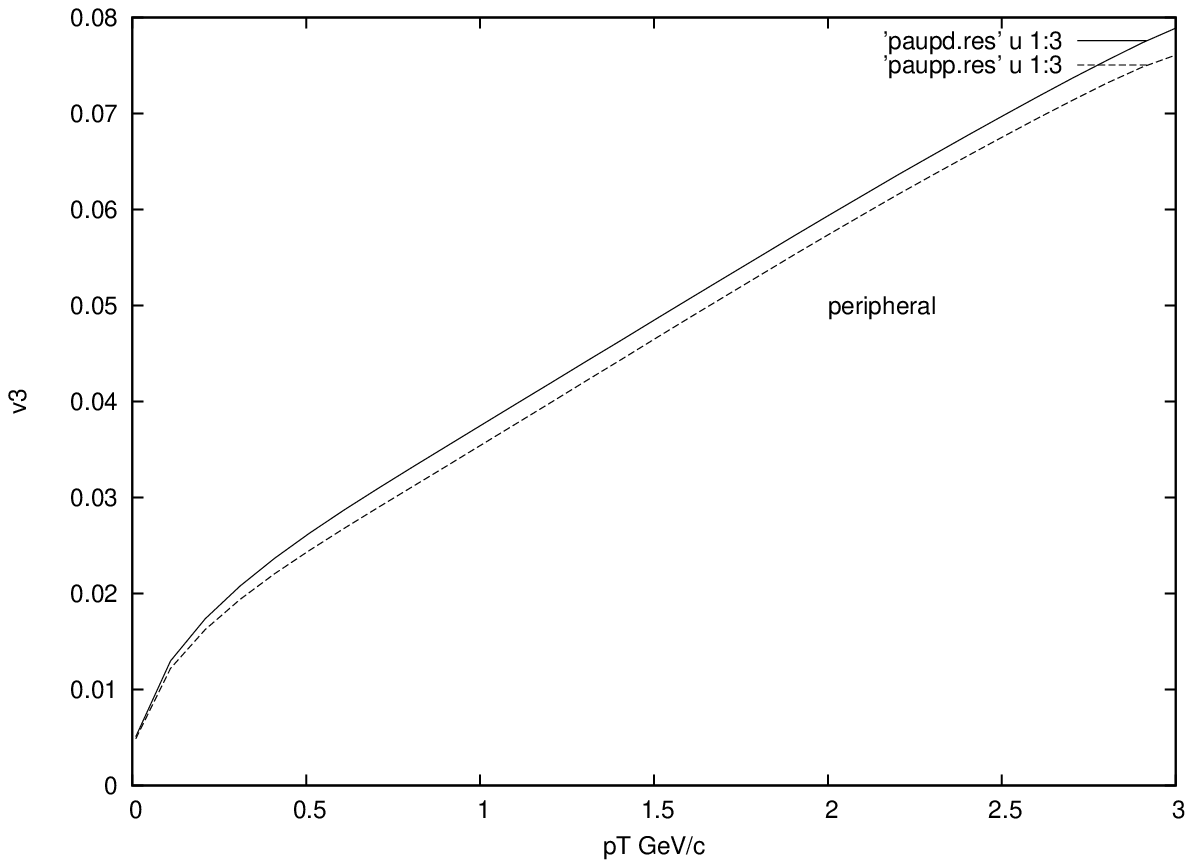}
\caption{The calculated flow coefficients $v_2$ (left panel) and $v_3$ (right panel) as function of transverse momenta $p_T$.
for d-Au  and p-Au  peripheral collisions at 200 GeV. For d-Au collisions $v_2$ corresponds to the lower curve and $v_3$
corresponds to the upper curve.}
\label{fig3}
\end{center}
\end{figure}

The results of these calculations for $v_2$ and $v_3$ are presented in Figs. \ref{fig1}-\ref{fig3} for $p-Au$ and $d-Au$
central, mid-central and peripheral collisions at 200 GeV. Comparison with the existing experimental data for central collisions
\cite{aidala} is shown in Fig. \ref{fig1}.

One has to take into account that these results were obtained by Monte-Carlo simulations with a number of runs limited by
our calculational possibilities. So we estimate their statistical error as around  5\%. Taking this into account we
observe that our model reproduces the experimental data on $v_2$ rather satisfactorily with some overshooting of the $d-Au$
data and some undershooting the $p-Au$ data. For $d-Au$ our results for $v_2$ resemble those of ~\cite{aidala} although for
$p-Au$ we get somewhat lower figures. As to $v_3$ or model gives results which substantially overshoot the data and in this respect
is similar to models based on the gluonic initial conditions with the subsequent hydrodynamical evolution  ~\cite{mace,mace1,schenke}.
As pointed out in ~\cite{mace1} a possible reason for this discrepancy may be related to the fact that unlike $v_2$ the triangularity $v_3$
is exclusively due  fluctuations and is therefore very sensitive to eventual hadronization. In our model this process is treated in a
simplified way based on the well-known parton-hadron duality. We hope that a more detailed study of this stage may improve the situation.

\section{Conclusions}
An immediate consequence of our results is that they correctly describe the relative order of $v_2$ at central collisions  for
$d-Au$ and $p-Au$ events in spite of doubts expressed in \cite{aidala} and based on the comparison of the number of emitting sources
in these two systems. In our case the number of the emitting sources (strings) is naturally larger in $d-A$ collisions than in
$p-A$ collisions (roughly twice). However they do communicate contrary to the comments in \cite{aidala} by means of the commonly created
gluonic field. As a result $v_2$ for central $d-A$ are nearly 50 \% larger than for central $p-A$ collisions.

It is interesting Figs. \ref{fig2} and \ref{fig3} show that  this difference is diminishing with centrality and for
peripheral collisions practically vanishes. This circumstance seems to be natural from pure geometrical
considerations. Obviously in  peripheral collisions the probability to find both of the
projectile nucleons interacting with the target is minimal (if one of them interacts the other will be mostly located
outside the nucleus). So the picture will not be different from $p-Au$ collisions. The small differences in Fig. \ref{fig3} should not
be taken seriously due to errors involved in our Monte-Carlo simulations.
It would be desirable to study this effect in the experiment.

As mentioned the triangularity $v_3$ comes out considerably larger for both $p-Au$ and $d-Au$ at central collisions.
In this respect our model gives results similar to the models based on the hydrodynamical evolution of the initial gluon density.
We hope that some improvement of our model taking into account fluctuations in the parton-hadron conversion may improve the
results for $v_3$. We are working on this problem.

\section{Acknowledgements}
M.A.B. appreciates hospitality and financial support of the University of Santiago de Compostela, Spain.
C.P. thanks the grant Maria de Maeztu Unit of Excelence of Spain and the support
Xunta de Galicia. This work was partially done under the project EPA 2017-83814-P
of Ministerio Ciencia, Tecnologia y Universidades of Spain.

\end{document}